\documentclass[12pt,twoside,pointlessnumbers,nofootinbib,smallheadings]{revtex4-1}
\usepackage{amsmath}
\usepackage{graphics}
\usepackage{slashed}
\usepackage{graphicx}
\usepackage{indentfirst}
\usepackage{amssymb}

\begin{document}

\title{ Anomalous Higgs-top Coupling Pollution on Triple Higgs Coupling Extraction at Future High-Luminosity Electron-Positron Collider}

\author{Chen Shen $^{1,}$}
\email{chenshenphy@pku.edu.cn}

\author{Shou-hua Zhu $^{1,2,3,}$}
\email{shzhu@pku.edu.cn}
\affiliation{
$ ^1$Institute of Theoretical Physics $\&$ State Key Laboratory of
Nuclear Physics and Technology, Peking University, Beijing 100871,
China}
\affiliation{$ ^2$Collaborative Innovation Center of Quantum Matter, Beijing, China }
\affiliation{$ ^3$Center for High Energy Physics, Peking University,
Beijing 100871, China}


\begin{abstract}

One of the most challenging tasks for future high-luminosity electron-positron colliders is to extract Higgs triple coupling. It was proposed that this can be
carried out via the precisely measuring the cross section of ZH associated production up to $0.4\%$. In this paper, we example the possible heavy pollution
from Higgs-top anomalous coupling. Our numerical results show that the pollution is  small for $\sqrt{s}_{e^+e^-}= 240 GeV$. However for the higher energy collider, pollution
is sizable, which should be taken into account. We further explored the possibility to measure CP-violated Higgs-top coupling, via the forward-backward asymmetry $A_{FB}$ for
the process $e^+e^- \rightarrow ZH$. The asymmetry can reach $0.7\%$ which is comparable to the precision of cross section measurement.

\end{abstract}

\maketitle

\newpage

\section{Introduction}

A standard model (SM) like Higgs boson \cite{higgs} (denoted as H(125) in this paper) was discovered at the Large Hadron Collider (LHC) in 2012.
In order to test the SM and discover possible physics beyond the SM (BSM),
it is crucial to measure the Higgs Yukawa couplings and Higgs self couplings at the LHC and future high energy colliders.
In the first run of the LHC, the CMS and ATLAS has constrained $h\bar t t$ Yukawa coupling indirectly through the global fit,
with a precision of  20$\%$ and  30$\%$  respectively \cite{Khachatryan:2014jba,a}.
With 300/fb, Yukawa couplings will be measured up to 23$\%$, 13$\%$  and 14$\%$ for $h\bar b b$, $h\tau^+ \tau^-$ and $h\bar t t$
respectively \cite{Peskin:2012we}.
It was also proposed to measure the top Yukawa coupling via the associated Higgs boson production with a single top quark
\cite{Barger:2009ky,Ellis:2013yxa,Biswas:2012bd,Biswas:2013xva,Farina:2012xp,Englert:2014pja,Chang:2014rfa}.
 The Higgs self coupling can be measured up to $50\%$ at the LHC with 300/fb \cite{Peskin:2012we}.
There are extensive studies on measuring anomalous  triple Higgs coupling directly at the LHC \cite{Baur:2002rb,Baur:2002qd,Baur:2003gp,Dolan:2012rv,Baglio:2012np} and
future electron-positron colliders  \cite{Baer:2013cma,Asner:2013psa}.

For the future high-luminosity electron-positron colliders, it is proposed to measure
the Higgs self coupling up to $28\%$ for $\sqrt{s_{e^+e^-}}=$ 240 GeV under the model-dependent assumption that only the Higgs self coupling is modified \cite{McCullough:2013rea} .
The  precision of Higgs self coupling  can only be reached based on the precisely measured cross section of ZH associated production  up to $0.4\%$ \cite{Gomez-Ceballos:2013zzn}.
 Entering $e^+e^- \rightarrow ZH$ via loops, the  triple Higgs coupling will be possibly polluted heavily by other anomalous couplings,
 and among them the dominant one is the $h-Z-Z$ coupling which appears even at tree-level.
 The first run results of LHC shows that the HVV couplings including $h-Z-Z$ coupling are consistent with those in the SM \cite{Khachatryan:2014jba,a}.
 The Higgs-top coupling contributes to the process
$e^+e^- \rightarrow ZH$ via loops and is potentially important for triple Higgs coupling extraction.
Actually the full one-loop correction to $e^+e^- \rightarrow ZH$ in the SM was calculated about two decades ago \cite{Kniehl:1991hk,Denner:1992bc,Fleischer:1982af,Denner:1991ue}.
In this paper we will focus on the anomalous Higgs-top coupling, especially its effects on the extraction of  triple Higgs coupling.

This paper is arranged as following. In section II, we estimate the deviation of the cross section for the process $e^+e^- \rightarrow ZH$ arising from anomalous Higgs-top coupling,
 and compare to that from triple Higgs coupling. In section III, we explore how to measure CP-violated Higgs-top coupling via the forward-backward asymmetry $A_{FB}$ for
the process $e^+e^- \rightarrow ZH$. The last section contains our conclusion and discussion.

\section{ Pollution from Higgs-Top anomalous Coupling}

\begin{figure}[!htbp]
\centering
\includegraphics[width=2.3in,totalheight=1.2in]{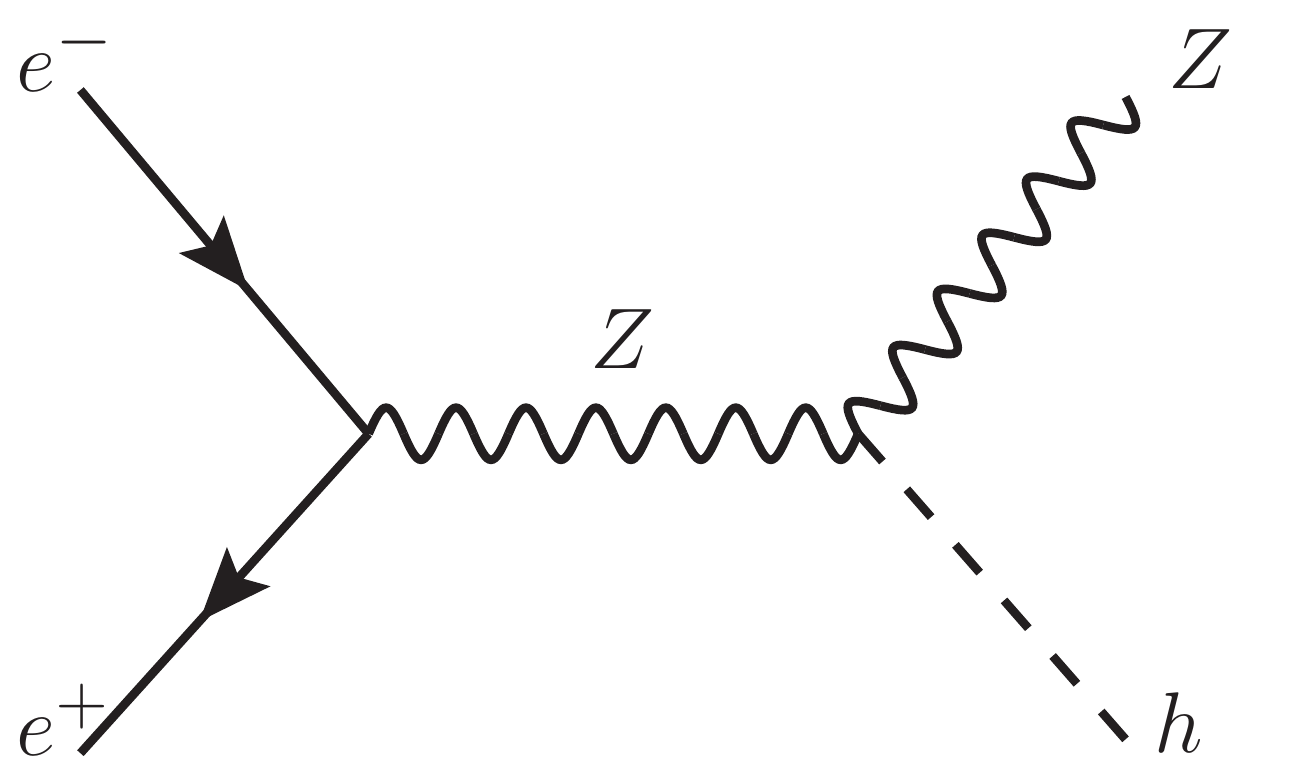}
\caption{Feyman diagram at tree-level for the process $e^+ e^- \rightarrow Zh$}
\label{fig1}
\end{figure}

In the SM, the process $e^+e^- \rightarrow ZH$ occurs at tree level and the Feynman diagram is shown in Fig. \ref{fig1}.

\begin{figure}[!htbp]
\includegraphics[width=2.3in,totalheight=1.2in]{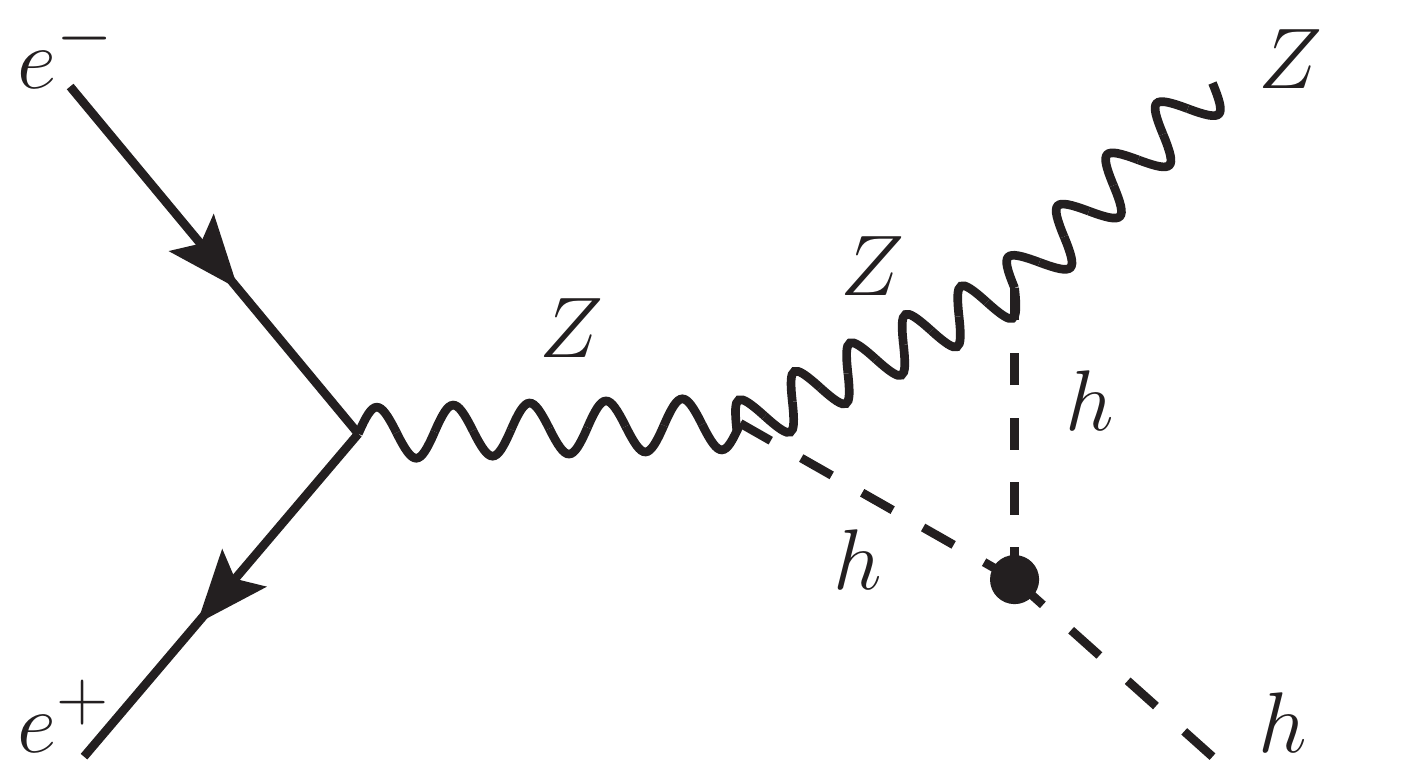}
\includegraphics[width=2.3in,totalheight=1.2in]{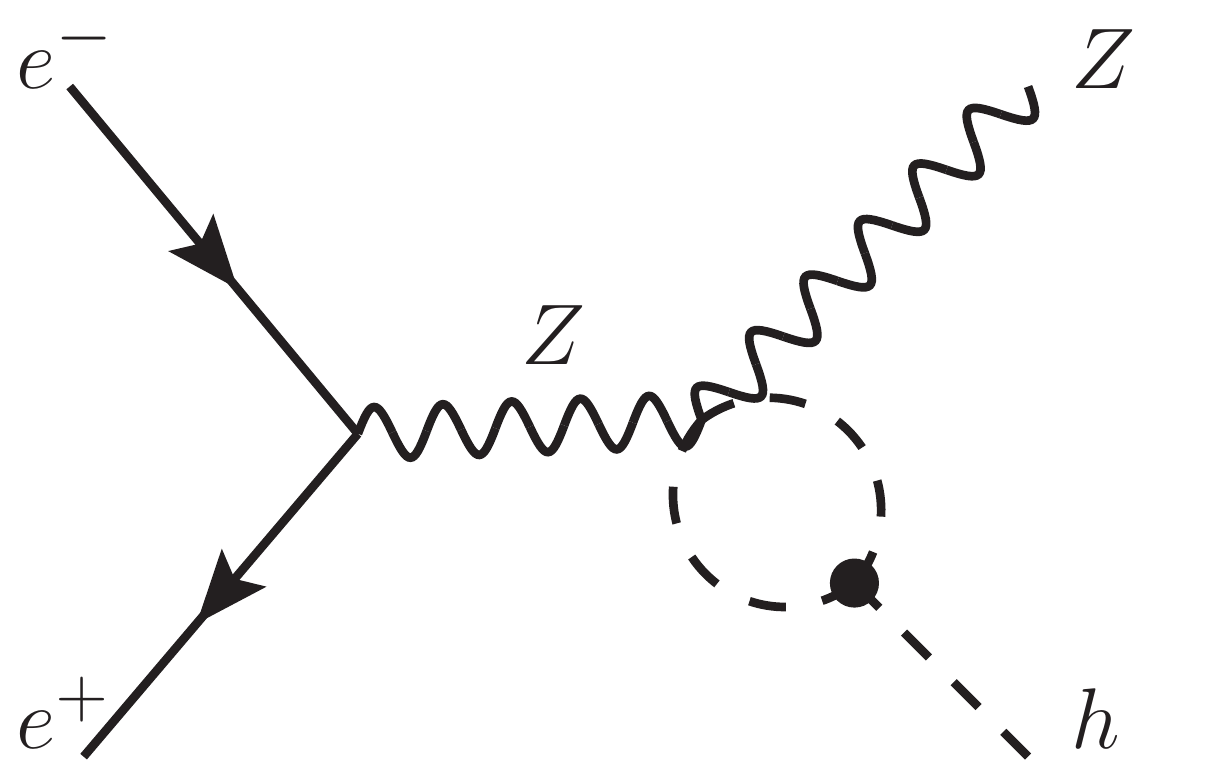}
\caption{Feynman diagram containing the anomalous $3h$ coupling, depicted as the black dot, at one-loop level for the process $e^+ e^- \rightarrow Zh$. }
\label{fig2}
\end{figure}
In order to measure the triple Higgs coupling, one way is to produce the Higgs pair, provided that the center of mass energy of $e^+e^-$ is high enough via $ e^+e^- \rightarrow HHZ$ or
$ e^+e^- \rightarrow HH \nu \bar \nu$ \cite{Levy:2015fva}. For such processes, the cross sections are notorious small.
High energy and high luminosity are both required. Another way to measure the
triple Higgs coupling is via the virtual effects which are shown in Fig. \ref{fig2}.
The capacity of measuring triple Higgs has been estimated by ref \cite{McCullough:2013rea}. For completeness we recalculate the analytical result for
$$\delta_{\sigma}\equiv \frac{\Delta \sigma}{\sigma}=\frac{\sigma_{\delta_h \ne 0}-\sigma_{\delta_h = 0}}{\sigma_{\delta_h = 0}}$$ from the triple Higgs coupling
$C_{SM}(1+ \delta_h) H H H=-3i \frac{m^2_h}{v}(1+ \delta_h) H H H $ as
\begin{equation}
\begin{split}
  \delta_{\sigma}(3h) =\frac{3 \alpha m_h^2 \delta_h} {16 \pi \beta c_w^2 s_w^2 m_z^2}
                    Re\Big[&2\rho \Big (C_1(m_h^2)+C_{11}(m_h^2)+C_{12}(m_h^2) \Big)\\
                    & -\beta \Big (B_0-4 C_{00}(m_h^2)+4 m_z^2C_0(m_h^2)+3m_h^2 B'_0 \Big ) \Big],
\end{split}
\end{equation}
where $\delta_h=0$ corresponds to the case in the SM.
Here $$ \beta=m_h^4-2 m_h^2 (m_z^2+s)+m_z^4+10 m_z^2 s+s^2,$$
$$ \rho=(m_h^2-m_z^2-s)\left((m_h-m_z)^2-s\right)\left((m_h+m_z)^2-s\right)$$
The definition of the one loop scalar functions B, C etc. can be found in Ref. \cite{Ellis:2007qk} and
$ C_0(m_h^2)=C_0(m_h^2,m_z^2,s,m_h^2,m_h^2,m_z^2)$,
$ C(m_h^2)=C(m_h^2,s,m_z^2,m_h^2,m_h^2,m_z^2)$,
$ B_0=B_0(m_h^2,m_h^2,m_h^2)$,
$ B'_0=\frac{\partial B_0}{\partial p^2}\bigg|_{p^2=m_h^2}$.

\begin{figure}[!htbp]
\includegraphics[width=2.3in,totalheight=1.2in]{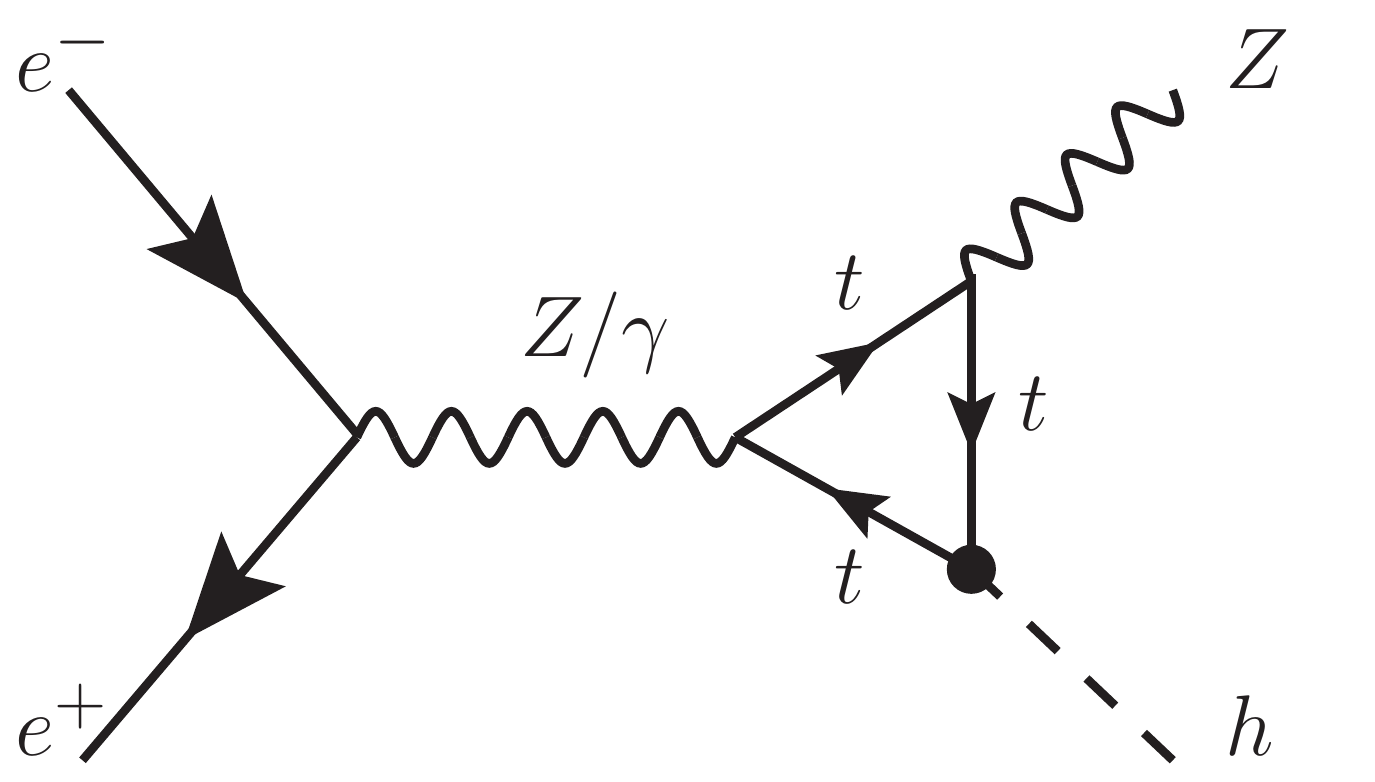}
\includegraphics[width=2.3in,totalheight=1.2in]{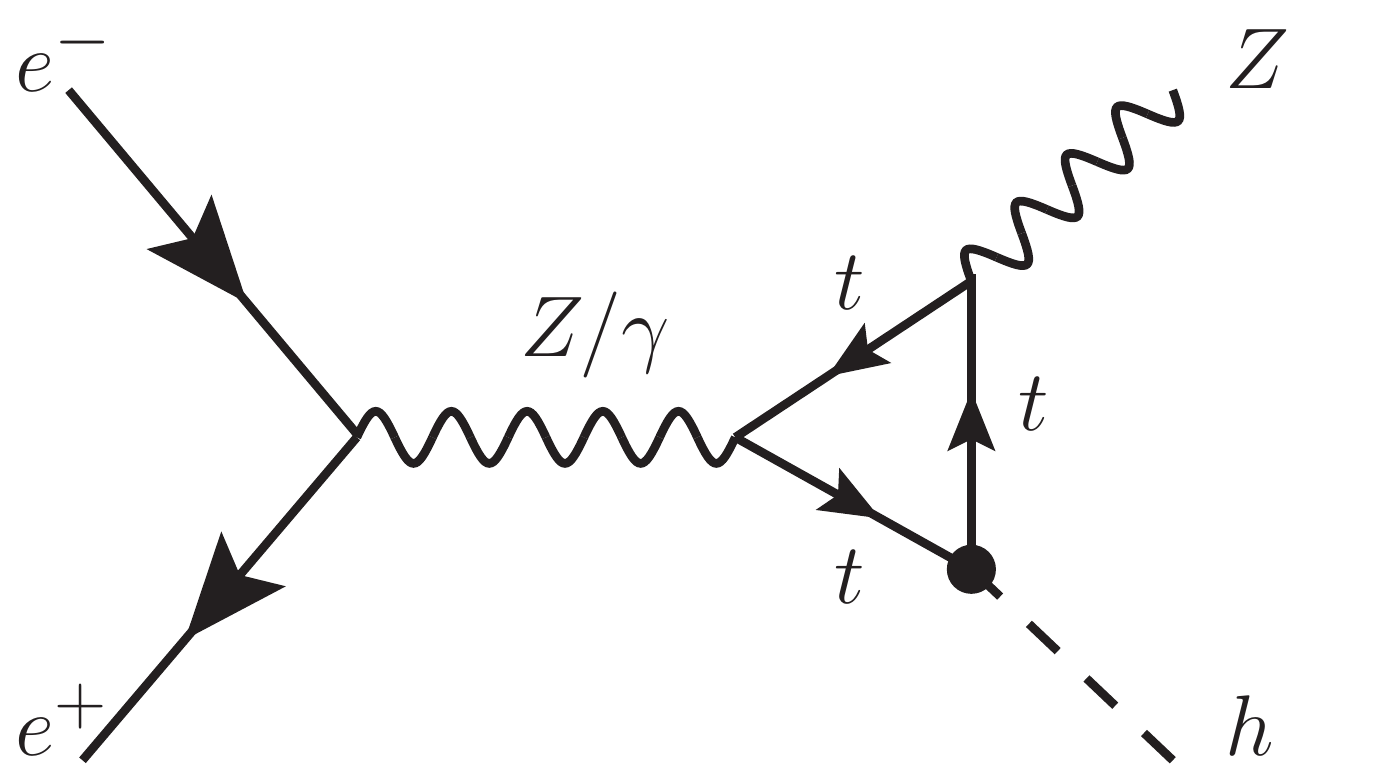}
\caption{Feynman diagram containing the anomalous $h\bar t t$ coupling, depicted as the black dot, at one-loop level for the process $e^+ e^- \rightarrow Zh$.}
\label{fig3}
\end{figure}
In this paper we will calculate the contributions from Higgs-top coupling which are shown in Fig. \ref{fig3}
\footnote{In fact,the contributions from $Z/\gamma-H$ bubble transition diagrams are zero.}. The Higgs-top coupling can be parameterized as
$$
C_{SM}(1+ \delta_t) H \bar t t=-i \frac{m_t}{v}(1+ \delta_t) H \bar t t,
$$
where $\delta_t=0$ corresponds to the case in the SM.

The analytical results can be written as
\begin{equation}
 \begin{split}
 &\delta_{\sigma}(htt)\\
  =&-\frac{4 N_c\alpha m_t^2 (s-m_z^2) v_1 v_2 \delta_t}{3 \pi s \beta m_z^2 (v_1^2+a_1^2)}Re\Big[\beta \Big (B_0(m_h^2)-4 C_{00}(m_t^2) \Big)\\
  &-2 \rho \Big(C_1(m_t^2)+C_{11}(m_t^2)+C_{12}(m_t^2) \Big)-6 m_z^2 s \big(s+m_z^2-m_h^2 \big ) C_0(m_t^2)\Big] \\
                       &+\frac{N_c \alpha m_t^2 \delta_t}{\pi c_w^2 s_w^2 m_z^2 \beta}  Re\Big[ \beta \Big( 2 \big (v_2^2+a_2^2 \big )\big(B_0(m_h^2)- 4 C_{00}(m_t^2)\big) +2 a_2^2 \big (B_0(s)+B_0(m_z^2) \big) \Big)  \\
                        &-\rho\Big( 4 \big (v_2^2+a_2^2\big )\big (C_1(m_t^2)+C_{11}(m_t^2)+C_{12}(m_t^2)\big)+2a_2^2 C_2(m_t^2)  \Big)\\
                        &+\Big(\big (v_2^2+a_2^2 \big) \big((4 m_t^2- m_z^2-s) \beta - \rho \big) +\big (v_2^2-a_2^2 \big) \big (m_h^2-4 m_t^2 \big ) \beta \Big) C_0(m_t^2)  \Big]\\
                      &+\frac{N_c\alpha m_t^2 \delta_t}{4 \pi c_w^2 s_w^2 m_z^2}Re\Big[-B_0(m_h^2)+\big(4 m_t^2- m_h^2 \big) B'_0(m_h^2)\Big]
 \end{split}
 \label{eq2}
\end{equation}
In Eq. (\ref{eq2}), the first/second/third terms are from the contributions of the diagram with photon propagator/Z boson propagator/the counter term of ZZH vertex, respectively.
Here
$ \alpha = \frac{e^2}{4\pi}$,
$N_c=3$,
$ v_1=-\frac{1}{4}+s_w^2$,
$ a_1=\frac{1}{4}$,
$ v_2=\frac{1}{4}-\frac{2}{3}s_w^2$,
$ a_2=-\frac{1}{4}$,
$ C_0(m_t^2)=C_0(m_h^2,m_z^2,s,m_t^2,m_t^2,m_t^2)$,
$ C(m_t^2)=C(m_h^2,s,m_z^2,m_t^2,m_t^2,m_t^2)$,
$ B_0(m_h^2)=B_0(m_h^2,m_t^2,m_t^2)$,
$ B_0(m_z^2)=B_0(m_z^2,m_t^2,m_t^2)$,
$ B_0(s)=B_0(s,m_t^2,m_t^2)$.

\begin{figure}[!htbp]

\includegraphics[width=0.3\textwidth]{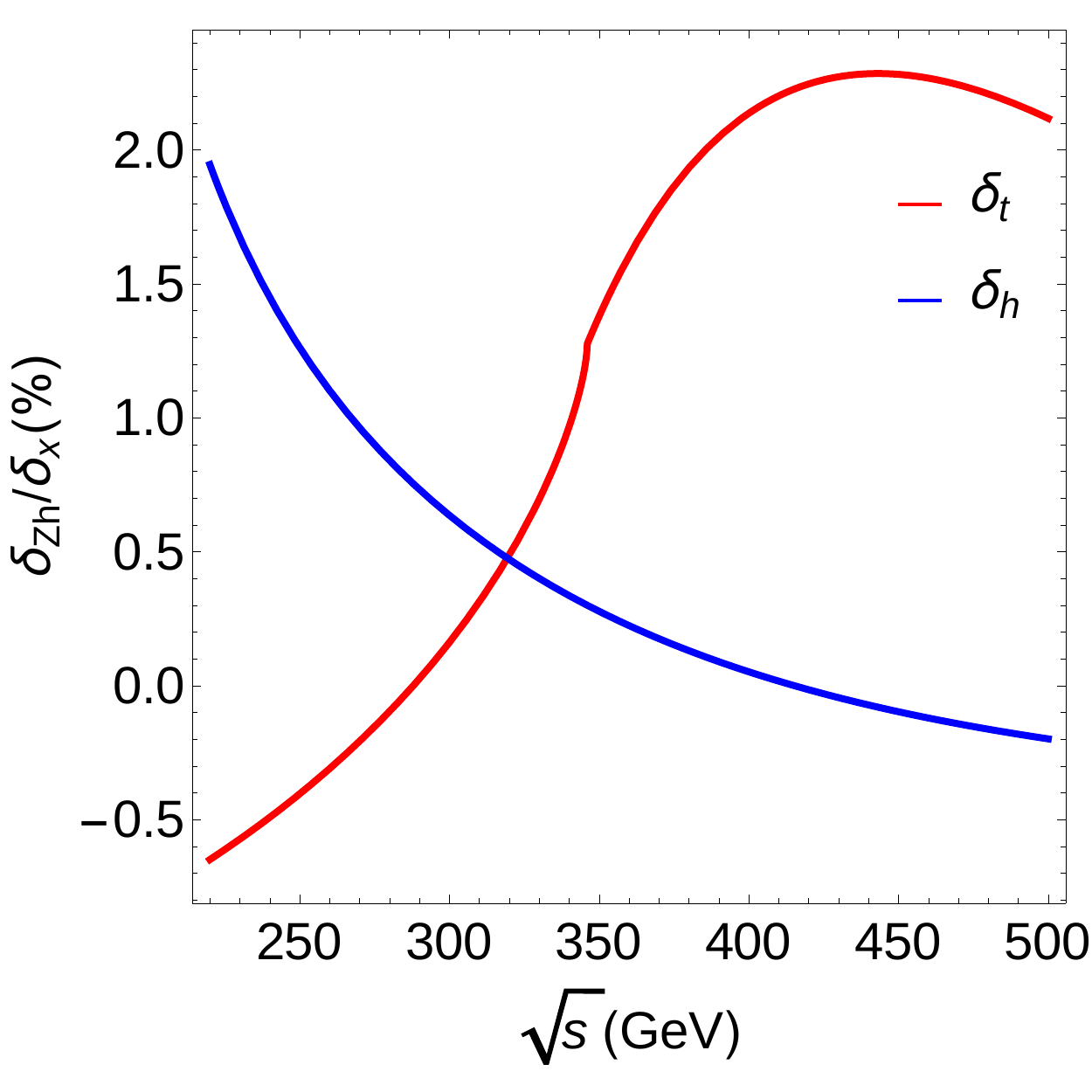}
\caption{ Relative correction $\delta_\sigma$  due to anomalous $h\bar{t}t$-coupling $\delta_t$ (red)
         and anomalous triple Higgs coupling $\delta_h$ (blue), as a function of the $e^+ e^-$ center-of-mass (c.m.) energy from 220 GeV to 500 GeV. Note that the precision of
         relative correction can reach $0.4\%$ for high luminosity $e^+ e^-$ colliders. }
\label{fig4}
\end{figure}

We use LoopTools \cite{Hahn:1998yk} to do the scalar integral for different  c.m. energies.
In Fig. \ref{fig4}, we show the deviation of cross section arising from $\delta_t$ and  $\delta_h$ as a function of $\sqrt{s}_{e^+ e^-}$.
Several numerical results for the typical c.m. energy are
\begin{equation}
 \delta_\sigma^{240,350,400,500}=1.45,0.27,0.05,-0.19\times \delta_h \%
\end{equation}
 \begin{equation}
 \delta_\sigma^{240,350,400,500}=-0.49,1.38,2.14,2.12\times \delta_{t} \%
\end{equation}

The figures show that the behavior for $\delta_t$ and $\delta_h$ is opposite. At low energy end, the relative correction $\delta_\sigma$ happen to be dominant by $\delta_h$, on
the contrary for the high energy end, the $\delta_\sigma$ arising from anomalous Higgs-top coupling can't be neglected.
For the proposed collider of Circular Electron-Positron Collider with $\sqrt{s}_{e^+e^-} \simeq 240$ GeV, the extraction
of triple Higgs coupling is  polluted by Higgs-top coupling. For the International Linear Collider with option of high energy,
the pollution from Higgs-top coupling must be taken into account.

\section{ Measuring CP-violated Higgs-Top Coupling}

Though the newly discovered Higgs boson H(125) is SM-like, it does not exclude the possibility that H(125) is CP mixing state. As emphasized by \cite{zhu,Mao:2014oya} that
CP spontaneously broken \cite{Lee} may be closely related to the lightness of the H(125). In fact, current measurements are insensitive to the mixing, especially for
H decaying into gauge bosons since the CP violation usually entering the couplings via loops.

In this paper we parameterize the CP violation through
$$
C_{SM} H\left(1+\delta_t +i \delta_a \gamma_5\right)=-i \frac{m_t}{v} H\left(1+\delta_t +i \delta_a \gamma_5\right).
$$
Indirect constraints on $\delta_t$ and $\delta_a$ at the LHC have been studied in \cite{Ellis:2013yxa}.
At the 68\% CL  the allowed region for ($1+\delta_t$ , $\delta_a$) is a crescent with apex close to the SM point(1,0) \cite{Ellis:2013yxa}.
The parameter space close to the SM point, namely $\delta_t \rightarrow 0$ and $\delta_a \rightarrow 0$ is allowed. At the same time, the parameter space with
both non-zero $\delta_t,  \delta_a$ is also allowed. In fact, it is quite challenging for LHC to completely exclude the latter case via the indirect method.
On the contrary, based on the last section analysis, the cross section deviation depends only on $\delta_t$ but not $\delta_a$. This point will be made clear below.  Therefore
it is important to explore the method to measure the $\delta_a$ at electron-positron collider.

The analytical results for the differential cross section arising from $\delta_a$ can be written as
\begin{equation}
 \begin{split}
 &\frac{1}{\delta_a} \frac{d\sigma}{d cos\alpha}\\
 =&\frac{32 N_c a_1 m_t^2 \pi \alpha^3 cos\alpha \sqrt{\left((m_h-m_z)^2-s\right) \left((m_h+m_z)^2-s\right)}}{c^4_w s^4_w \left(m_z^2-s\right)}\\
 &Im\Big[\frac{1}{3} v_2  C_0(m_t^2)
   +\frac{ s}{c_w^2 s_w^2 \left(m_z^2-s\right)}  v_1
\big((v_2^2+a_2^2) C_0(m_t^2)+2 a_2^2 C_2(m_t^2)\big)\Big]
 \end{split}
\end{equation}
Here $cos\alpha$ is the angle between the momentum of the electron and the Z boson. The differential cross section is  proportional to $cos\alpha$, which is
due to the term $\varepsilon_{\mu \nu \rho \lambda}  \varepsilon^{\mu \nu \alpha \beta}  p_2^{\rho}  p_1^{\lambda}  k_{1\alpha}  k_{2\beta}$
where $p_1$ $p_2$ are the momentum of electron and positron and $k_1$ $k_2$ are the momentum of Higgs and Z. Another critical requirement for non-vanishing contribution
to the differential cross section
is that there should be imaginary part from top loops. This requires that the $\sqrt{s}_{e^+e^-}$ must be great than $2 m_t$.

It is obvious that the CP-odd contributions to the total cross section is zero.
In order to show the different contributions from $\delta_t$ and $\delta_a$ respectively, we plot the normalized differential cross sections for several $\sqrt{s}_{e^+e^-}$
and set the corresponding parameter $\delta_t$ or $\delta_a$ equal to 1.
\begin{figure}[!htbp]
\includegraphics[width=0.5\textwidth]{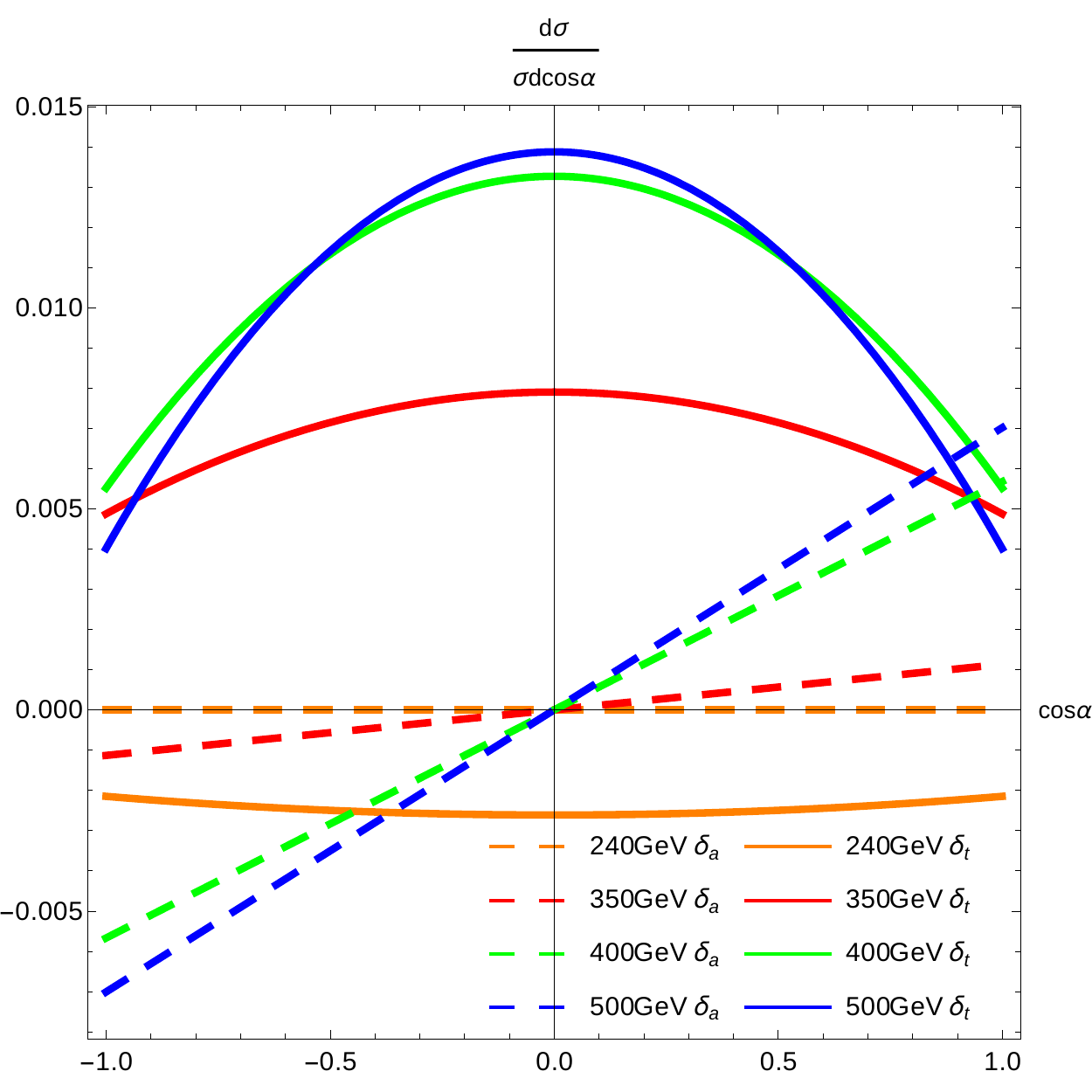}
\caption{Differential scattering cross section as a function of the scattering angle with $\sqrt{s}=240GeV$(orange),$350GeV$(red),$400GeV$(green),$500GeV$(blue).
And solid/dashed lines stand for the contributions
 from  $\delta_t$/$\delta_a$ respectively. }
\label{6}
\end{figure}
From the  figure, it is quite clear that the differential cross sections arising from $\delta_t$ are symmetric and anti-symmetric from $\delta_a$.
For $\sqrt{s}=240GeV$, the contribution from $\delta_a$ is zero because there is no imaginary part of $ C_0(m_t^2)$.
When $\sqrt{s}_{e^+e^-} > 2 m_t$ there are nonzero contributions from $\delta_a$ as expected.

In order to gauge the forward-backward asymmetry, we introduce
$$
A_{FB} \equiv \frac{\int_{0}^1 d\cos \alpha \frac{d\sigma}{d\cos\alpha}  -  \int_{-1}^0 d\cos \alpha \frac{d\sigma}{d\cos\alpha} }{\sigma_{tot} }
$$

\begin{figure}[!htbp]
\includegraphics[width=0.4\textwidth]{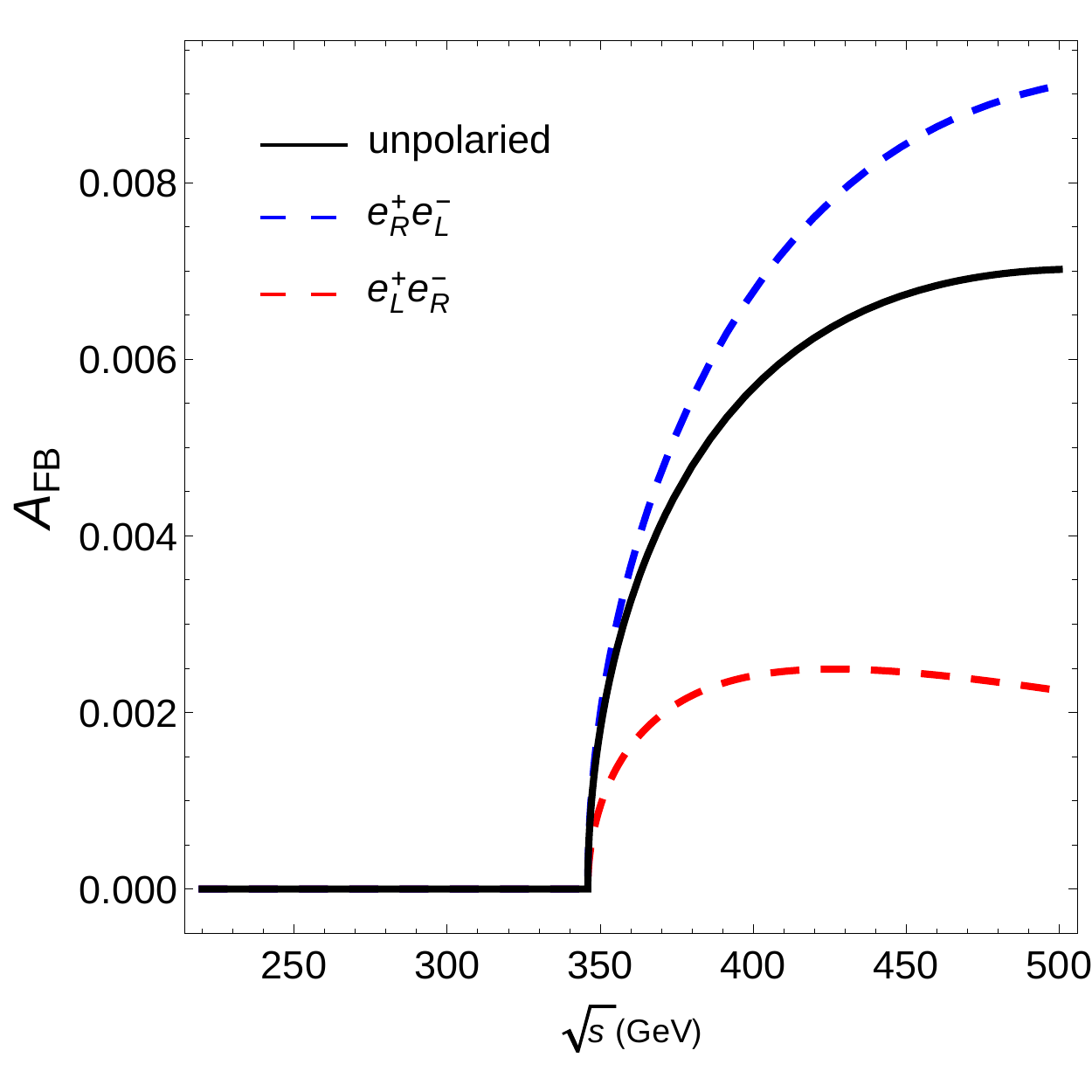}
\caption{  $A_{FB}$ as a function of $\sqrt{s}_{e^+e^-}$ from $220GeV$ to $500GeV$ for polarized or unpolarized electron/positron beams.
The black line, blue dashed and red dashed seperately
correspond to  unpolarized electron/positron beams,$e_R^+ e_L^-$ and $e_L^+ e_R^-$ polarizations.}
\label{fig9}
\end{figure}

In Fig.  \ref{fig9}, we plot $A_{FB}$ as a function of $\sqrt{s}_{e^+e^-}$ with $\delta_a=1$ for polarized and unpolarized electron/positron beam.

From the figure we can see that the asymmetry can reach $0.7 \%$ for $\sqrt{s}_{e^+e^-}$. Such precision is comparable to that of cross section measurement.
It seems that the high luminosity collider is necessary.

\section{Conclusion and discussion}

In this paper, we explore the Higgs-top anomalous coupling pollution to the extraction of Higgs self coupling via precisely measuring cross section of $e^+e^- \rightarrow ZH$.
The important conclusion is that the pollution is small for the $\sqrt{s}_{e^+e^-} =240$ GeV, but can be sizable for higher energy collider.
The contributions to total cross section from Higgs-top CP-odd coupling is vanishing, while such interaction can
 be scrutinized via forward-backward asymmetry for $\sqrt{s}_{e^+e^-}$ greater than $2 m_t$.

\section*{Acknowledgement}

 We would like to thank Shao Long Chen , Gang Li and Pengfei Yin for the useful discussions. This work was supported in part by the Natural Science Foundation
 of China (Nos. 11135003 and 11375014).

\end{document}